\documentclass[twocolumn]{pasj00}
\usepackage{color}
\SetRunningHead{T. Sonoi and H. Shibahashi}
{Vibrational Instability of Population III Stars}
\Received{2011 April 28}
\Accepted{2011 August 2}
\Published{}
\begin{document}
\title{Fully Nonadiabatic Analysis of \\
Vibrational Instability of Population III Stars \\
 due to the $\varepsilon$-Mechanism}
\author{Takafumi \textsc{Sonoi} and Hiromoto \textsc{Shibahashi}}
\affil{ Department of Astronomy, School of Science, The University of Tokyo  \\
7-3-1 Hongo, Bunkyo-ku, Tokyo 113-0033 }
\email{sonoi@astron.s.u-tokyo.ac.jp, shibahashi@astron.s.u-tokyo.ac.jp}
\maketitle

\begin{abstract}
A linear nonadiabatic analysis of the vibrational stability of population III main-sequence stars was carried out. It was demonstrated that, in the case of massive stars with $M \gtrsim 5\,\Mo$, helium burning (triple alpha reaction) starts during the main-sequence stage and produces $^{12}{\rm C}$, leading to the activation of a part of the CNO-cycle. It was found that, despite of that, those stars with $M\lesssim 13\Mo$ become unstable against the dipole g$_1$- and g$_2$-modes due to the $\varepsilon$-mechanism, during the early evolutionary phase at which the pp-chain is still the dominant nuclear energy source. The instability due to the $\varepsilon$-mechanism occurs against g-modes having a large amplitude in the off-centered $^3$He accumulation shell in the deep interior, and the growth time is much shorter than the evolutionary timescale. 
This instability is therefore likely to induce mixing in the stellar interior, and have a significant influence on the evolution of these stars.
\end{abstract}

\section{Introduction}
\label{sec:1} 
The structure and evolution of the first-generation stars (so-called population III stars) born in the very early universe with no, or little, heavy elements are considerably different from those of stars born much later with heavy elements (e.g.,\cite{Weiss2000},  \cite{Marigo2001}, \cite{Suda2007}). This is partly because the initial mass function is substantially different due to the lack of an efficient coolant at the star-formation stage, and also partly because the only possible way of starting of hydrogen burning in population III stars is the pp-chain, of which temperature dependence is weak, no matter how massive the star is. Hence, the temperature near the stellar center of intermediate and massive population III stars is much higher than in the case of the population I/II stars, and then these population III stars are, to a great extent, compact compared with the population I/II stars. Since the pp-chain is the main energy source, the temperature gradient in the nuclear burning core of metal-free population III stars is milder than in the case of population I stars with heavy elements. Consequently, the nuclear burning core is radiative, except for during the short period of the ZAMS stage. Furthermore, the high central temperature of population III stars leads to the occurrence of helium nuclear burning (triple alpha reaction) even during the main-sequence stage. On the other hand, no metallicity and the consequent low opacity make the envelope to be radiative, except for the surface layer. 

Since the structure of population III stars is considerably different in this way from that of stars with heavy elements, their stability nature might also be different. The present authors have already shown by using a quasi-adiabatic approximation that low-mass population III stars with  $M\lesssim 5\,\Mo$ become unstable during the main-sequence stage against the dipole low-order g-modes due to the $\varepsilon$-mechanism (\cite{Sonoi2011}, hereafter Paper I). The physical cause of this instability is explained as follows (\cite{Chris1974}; \cite{Boury1975}; \cite{Shibahashi1975}; see also Paper I): $^3{\rm He}$ accumulates in a shell just outside the stellar center in the process of the pp-chain, since the $^3$He-$^3$He reaction occurs so fast compared with the pp-reaction that $^3{\rm He}$ does not remain at the stellar center, while the $^3{\rm He}$-$^3{\rm He}$ reaction does not efficiently occur in the outer part of the nuclear burning core with a decrease in temperature. In such a situation, the high-temperature sensitivity of the $^3{\rm He}$-$^3{\rm He}$ reaction leads to a vibrational instability of g-modes having a large amplitude of temperature perturbation in such an off-centered $^3{\rm He}$ shell.  

In Paper I, when examining the vibrational stability of the g-modes, we adopted the quasi-adiabatic approximation, in which the work integral was evaluated by using the adiabatic eigenfunctions as well as adiabatic relations. This approximation is good enough in the bulk of the stellar interior, but ibecomes inappropriate in the very outer envelope where the thermal time scale is short and the nonadiabatic effect is large. To avoid this difficulty, the integration was terminated at a radius where the thermal timescale became equal to the oscillation period. As a consequence, the contribution of the $\kappa$-mechanism was not properly evaluated, though there appeared a tendency for a destabilization effect due to the $\kappa$-mechanism in some modes. In the present work, we carried out a fully nonadaiabatic analysis to see the additional effect of the $\kappa$-mechanism on the vibrational stability of the g-modes.  

In Paper I, we restricted ourselves to treat only low-mass stars, $M < 5\,M_\odot$, in which the CNO-cycle never starts to work at the early stage of core hydrogen-burning. In the case of more massive stars, even at the early stage of core hydrogen-burning, the central temperature becomes high enough for helium burning (triple alpha reaction) to start in addition to hydrogen burning due to the pp-chain. In such a case, the production of $^{12}{\rm C}$ by the triple alpha reaction leads to activation of, at least, a part of the CNO-cycle, and hence a careful treatment of the nuclear reaction networks is required. In the present work, we carried out stellar evolution calculations with nuclear networks, and extended the mass range of stars.  

The purpose of the present work is to examine the vibrational stability of population III stars more completely than Paper I by performing stability analysis by solving the fully nonadiabatic oscillations over a wide mass range of population III stars.

\section{Evolutionary equilibrium models of population III stars}
We adopted a recently developed, open source stellar evolution code, MESA (revision 3107, \cite{Paxton2011}), which is appropriate for the current purpose of providing state-of-the-art modules for equation of state, opacity, nuclear reaction rates, element diffusion data, and atmosphere boundary conditions. Evolutionary models of 1, 1.2, 1.4, 1.6, 2, 3, 5, 7, 10, 13, 16, and $20\,\Mo$ with the initial chemical composition $X_0=0.76$ and $Z=0$ were constructed from the pre-main sequence Hayashi phase.
Since this code carefully treats the change in the abundance of chemical elements, the evolution of population III stars is computed with sufficient precision even after the onset of the triple alpha reaction and the resultant activation of the CNO-cycle.
On the other hand, this code outputs, at default, physical quantities only below the photosphere; we therefore modified it to output an Eddington gray atmosphere as well by utilizing the MESA's module so that we could set the outer boundary conditions of the pulsation equations at an optically thin layer ({\it From revision 3372, MESA has given an option to output atmosphere models.}).

\begin{figure}
  \begin{center}
    \FigureFile(95mm,95mm){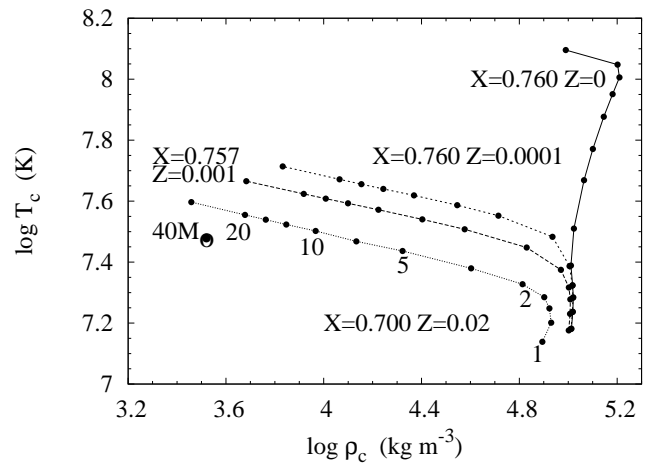}
  \end{center}
    \caption{Physical conditions at the center of 1, 1.2, 1.4, 1.6, 2, 3, 5, 7, 10, 13, 16, 20, 40$\Mo$ stars at the ZAMS stage with different metallicity. The abscissa is the density in units of ${\rm kg}\,{\rm m}^{-3}$, and the ordinate is the temperature.}
    \label{fig:01}
\end{figure}

\begin{figure}[h]
  \begin{center}
    \FigureFile(90mm,90mm){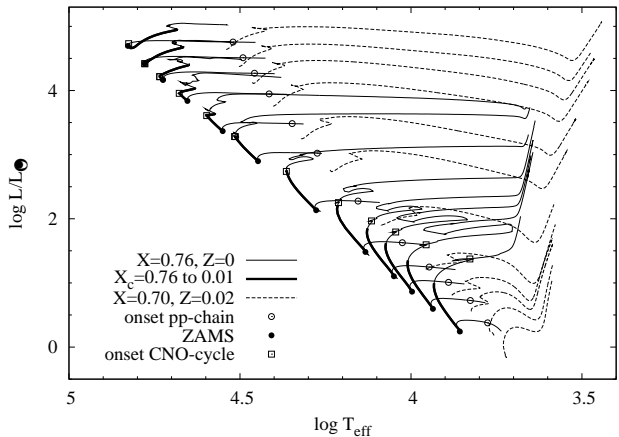}
  \end{center}
    \caption{Theoretical HR diagram showing the evolutionary tracks of 1.0, 1.2, 1.4, 1.6, 2.0, 3.0, 5.0, 7.0, 10.0, 13.0, 16.0, and $20.0\,\Mo$ stars with the initial composition $(X, Z)=(0.76, 0),(0.70,0.02)$. The bold lines on the $Z=0$ tracks indicate stages from $X_{\rm c}=0.76$ to $0.01$ ($X_{\rm c}$ is the central hydrogen abundunce), in which the stability analysis is performed in this study. In the case of $M \gtrsim 1.2\,\Mo$ stars, evolutionary tracks of metal-free stars are significantly different from those of $Z\ne 0$ because of suppression of the CNO-cycle.}
    \label{fig:02}
\end{figure}

A lack of heavy elements leads to low opacity, and thus the luminosity of population III stars is substantially higher than in the case of stars with heavy elements. This means that the population III stars are compact compared with population I stars; also, the central density and temperature are significantly higher than in the case of population I stars. This feature is clearly seen in figure \ref{fig:01}. A conspicuous difference between metal-free stars and stars with $Z\ne 0$ in this diagram is that the central density becomes higher and higher with an increase in the mass of the star, up to $M \simeq 16\,\Mo$, in the case of metal-free stars, while in the case of $Z\ne 0$ it turns to become lower with an increase of the mass of the stars, for $M\gtrsim 1.2\,\Mo$. This difference is caused by the fact that the energy source of population I/II stars turns to be the CNO-cycle for $M \gtrsim 1.2\,\Mo$ from the pp-chain, while in the case of metal-free stars the pp-chain remains as the only source, even in the mass range of $M \gtrsim 1.2\,\Mo$. While there appears to be a convective core in the case of metal-free stars with $1.2\Mo \lesssim M \lesssim 16\Mo$ at the ZAMS stage, and very close to it, despite that the nuclear reaction source is the pp-chain, due to the high-temperature gradient, the nuclear burning core soon becomes radiative in the case of lower mass. On the other hand, for stars with $M \gtrsim 16\,\Mo$, the central temperature becomes high enough for helium burning to occur at the ZAMS stage. As a consequence, the CNO-cycle is activated with the produced $^{12}{\rm C}$ at zero-age, and this process dominates over the pp-chain. This is the reason why the central density becomes lower with an increase of the mass of the star for $M \gtrsim 16\,\Mo$. The nuclear burning core of these stars is convective. Figure \ref{fig:02} shows the evolutionary tracks of the stars on the HR diagram. Since the nuclear reaction occurring in population III stars of $M\lesssim 16\,\Mo$ at the ZAMS stage is the pp-chain, they go toward higher temperature and higher luminosity on the HR diagram, like in the case of $M\lesssim 1.2\,\Mo$ of population I stars.

\begin{figure}[th]
  \begin{center}
    \FigureFile(90mm,90mm){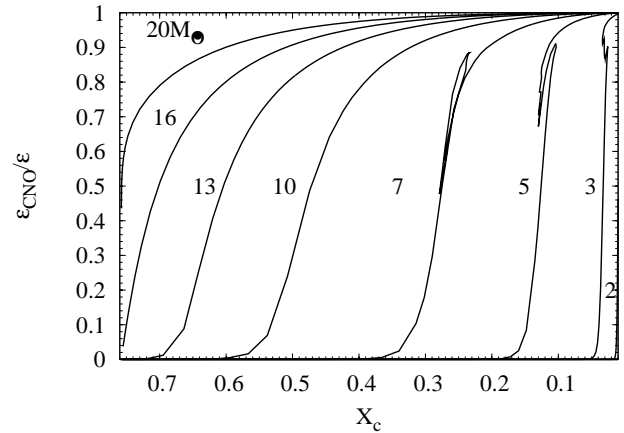}
    \FigureFile(90mm,90mm){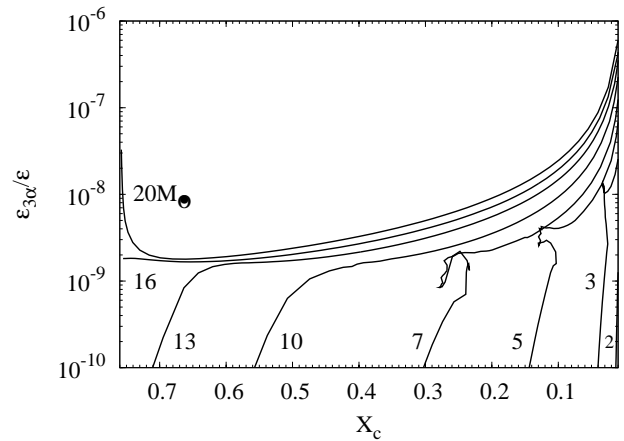}
  \end{center}
    \caption{Fraction of nuclear energy by the CNO-cycles and triple alpha reaction to the total  nuclear energy at the stellar center. The abscissa is the central hydrogen abundance, $X_{\rm c}$, which is an indicator of stellar evolution. This range is set to be from $X_{\rm c}=0.76$ to 0.01. For $\lesssim 1.6\,\Mo$, the nuclear energy is provided almost exclusively by the pp-chain. The contribution of the triple alpha reaction to the whole nuclear energy is negligible.}
    \label{fig:03}
\end{figure}

As can be seen in figure \ref{fig:03}, in the case of $2\,\Mo \lesssim M \lesssim 16\,\Mo$, though the contribution of helium burning to the total luminosity is negligibly smaller compared to  that of the pp-chain, it gradually becomes large enough during the main-sequence stage to produce $^{12}{\rm C}$, which activates the CNO-cycle of hydrogen burning \citep{Suda2007}. 
The CNO-cycle onset stage is marked by open squares in figure \ref{fig:02}. 
Since the temperature dependence of the CNO-cycle is much higher than that of the pp-chain, the CNO-cycle becomes to dominate over the pp-chain if once the CNO-cycle is activated. Also, the nuclear burning core becomes convective. Figure \ref{fig:03} shows the evolutionary change in the contribution of the CNO-cycle to the total nuclear energy generation for stars wherer $M \ge 2\,\Mo$. As can be seen in this figure, the contribution of the CNO-cycle induced by triple alpha reaction becomes more important with an increase of the mass of the star. For stars with $M\gtrsim 16\,\Mo$, the CNO-cycle works even at the zero-age. On the other hand, for stars with  $M \lesssim 2\,\Mo$, the nuclear energy is provided only by the pp-chain in most of core hydrogen-burning phase.        

The characteristic properties of a population III star as an oscillator have already been discussed in Paper I.
It is remarked here again that, for a given mass, the radius of the population III star is substantially less than in the case of population I stars, and hence the dynamical timescale, which gives a rough estimate of the oscillation periods, of population III stars is substantially shorter than that of population I stars.
Another important factor for the oscillation periods of g-modes is the profile of the Brunt-V\"ais\"al\"a frequency. In the case of low-mass population III stars, an absence of the convective envelope increases the g-mode frequencies in units of the dynamical timescale of the star. Combining these two factors, the oscillation periods of the g-modes in population III stars are shorter than in the case of population I stars, as long as the mass and the mode are fixed.

\section{Fully nonadiabatic analysis of vibrational stability}
Any eigenfunction of linear nonradial oscillations of stars is expressed in terms of the combination of a spatial function and a time-varying function. The latter is expressed by $\exp (i\sigma t)$, where $\sigma$ denotes the eigenfrequency. The former, the spatial part, is decomposed into a spherical harmonic function, which is a function of the colatitude and the azimuthal angle, and a radial function. Equations governing the radial functions lead to a set of sixth-order differential equations, of which coefficients are complex, including the terms of frequency $\sigma$. Together with the proper boundary conditions, this set of equations forms a complex eigenvalue problem with an eigenvalue $\sigma$. The real part of $\sigma$, $\sigma_{\mathrm{R}}$, represents the oscillation frequency, and the imaginary part of $\sigma$, $\sigma_{\mathrm{I}}$, gives the growth rate, or the damping rate, depending on its sign. We have newly developed a Henyey-type code to solve this eigenvalue problem, following the formulation by \citet{Unno1989}. 
 
For a stability analysis relevant to the $\varepsilon$-mechanism, it is important to evaluate the temperature and density dependence of nuclear energy generation on the oscillation time scale, $\varepsilon_T \equiv(\partial\ln\varepsilon/\partial\ln T)_\rho$ and $\varepsilon_\rho \equiv(\partial\ln\varepsilon/\partial\ln\rho)_T$. It should be noted here that we should adopt the dependences of the nuclear reaction through perturbations, and that they are different from those for the evolutionary time scale. For example, as for the pp-chain, the temperature dependence of energy generation all through the pp-chain in equilibrium is governed by the slowest reaction, $^1{\rm H}(^1{\rm H}, {\rm e}^+\nu_{\rm e})^2{\rm H}$, and $\varepsilon_T \simeq 4$; we should thus adopt the effective temperature dependence of the nuclear reaction through the perturbation, which is mainly governed by $^3{\rm He}(^3{\rm He}, 2^1{\rm H})^4{\rm He}$ and $\varepsilon_T \simeq 11$ (\cite{Dilke1972}; \cite{Boury1973}; \cite{Unno1975}; \cite{Unno1989}). We evaluated these effective dependences of the pp-chain and of the CNO-cycle through the perturbation, separately, and then averaged them with their contribution to the total nuclear energy generation to obtain the net values. 

The temperature and density dependence of the pp-chain were evaluated following \citet{Unno1989}. For the CNO-cycle we took only the CN-cycle into account, since the occurrence probability of the other processes of the CNO-cycle is very low ($\sim 10^{-4}$). The CN-cycle involves relatively slow $\beta$-decay, of which the lifetime could be comparable to the oscillation period. \citet{Kawaler1988} derived the formulation for the temperature and density dependence, while taking into account the phase lag between the creation and destruction of various reactants. He used an approximation assuming that the time scales of $\beta$-decays in the CN-cycle $^{13}{\rm N}({\rm e}^-,\nu_{\rm e})^{13}{\rm C}$ and $^{15}{\rm O}({\rm e}^-,\nu_{\rm e})^{15}{\rm N}$ are comparable to the oscillation period $\Pi$, while those of the other collisional reactions are much longer than $\Pi$.
However, since in population III stars the onset of the CNO-cycle occurs at extremely high temperature ($\sim 10^8$K), just after $^{12}{\rm C}$ is produced by the triple $\alpha$ reaction, the time scale of the collisional reactions can also be comparable to the oscillation period $\Pi$. Therefore, we evaluated the temperature and density dependences of CN-cycle without assuming that the collisional reactions are much slower than the oscillation. For the sake of simplicity, we ignored the branch $^{13}{\rm N}({\rm p},\gamma)^{14}{\rm O}({\rm e}^-,\nu_{\rm e})^{14}{\rm N}$, which must occur at $\sim 10^8\,{\rm K}$. 
The details of the evaluation of the density and temperature dependence of the CN-cycle are given in the Appendix.

The stability is directly determined by the imaginary part of $\sigma$. It should be noted here that the work integral is useful to see where the excitation and damping zones are located in the stellar interior. The work integral, $W$, is defined as the amount of energy that must be removed from the star in order for the star to oscillate strictly periodically, and it is related to the imaginary part of $\sigma_{\rm I}$ through
\begin{equation}
\sigma_{\rm I}=-{{1}\over{2}}{{W/E_W}\over{\Pi}},
\label{eq:01}
\end{equation}
where $\Pi=2\pi\sigma_{\rm R}^{-1}$ denotes the period and $E_W$ is the total energy of oscillation, which is twice the time average of the oscillation kinetic energy,
\begin{equation}
E_W={{\sigma_{\rm R}^2}\over{2}}\int_0^M |\mbox{\boldmath$\xi$}|^2\,dM_r,
\end{equation}
where $\mbox{\boldmath$\xi$}$ denotes the eigenfunction of the displacement. The work integral was evaluated in Paper I using the quasi-adiabatic approximation. In the present study, we carried  out fully nonadiabatic calculations, and evaluated the work integral by 
\begin{equation}
W(r) = -{{1}\over{4}}\int_0^{M_r} \Im\left({{\delta T^*}\over{T}} T\delta S\right)\, dM_r,
\end{equation}
where $\delta T$ and $\delta S$ denote the eigenfunction of the Lagrangian perturbation of temperature and that of entropy, respectively. $W(r)$ represents the work done on the overlying layer by a sphere of radius $r$; its surface value is $W$, used in equation (\ref{eq:01}).

\section{Instability due to $\varepsilon$-mechanism of the pp-chain}
As discussed in section \ref{sec:1}, the favorable modes for the $\varepsilon$-mechanism of $^3{\rm He}$-$^3{\rm He}$ reaction are the low-degree g-modes having a large amplitude in the off-centered $^3{\rm He}$ accumulation shell, which is near to the outer edge of the nuclear burning core. Therefore, we investigate the dipole $(l=1)$ and quadrupole $(l=2)$ modes. 

\begin{figure}[th]
  \begin{center}
 \FigureFile(90mm,90mm){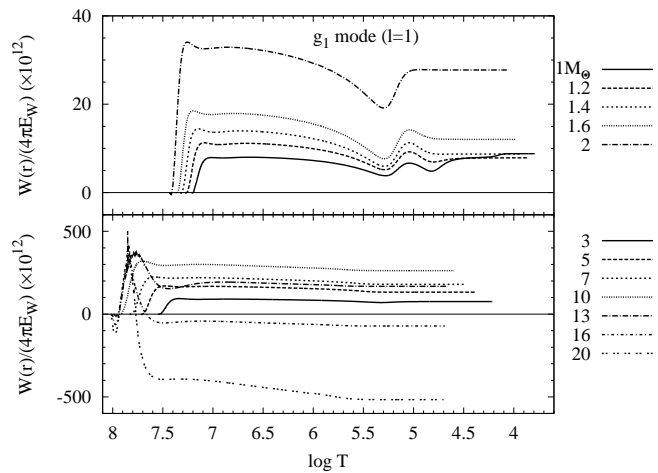}
  \end{center}
    \caption{Work integrals for diplole ($l=1$) g$_1$-mode in stars of 1-20\,$\Mo$. Equilibrium models correspond to the filled circles in the top panel of figure \ref{fig:05}.}
    \label{fig:04}
\end{figure}

\begin{figure}[!th]
  \begin{center}
    \FigureFile(90mm,90mm){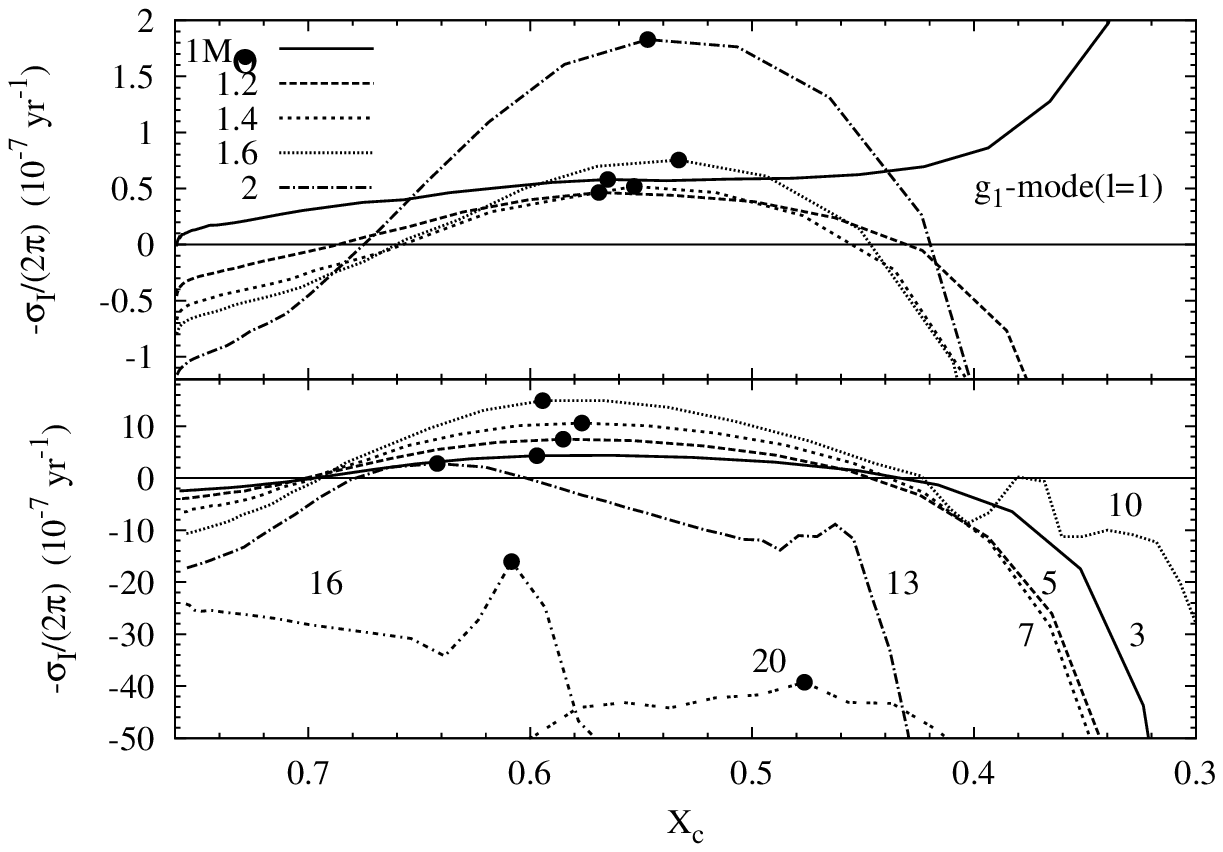}
    \FigureFile(90mm,90mm){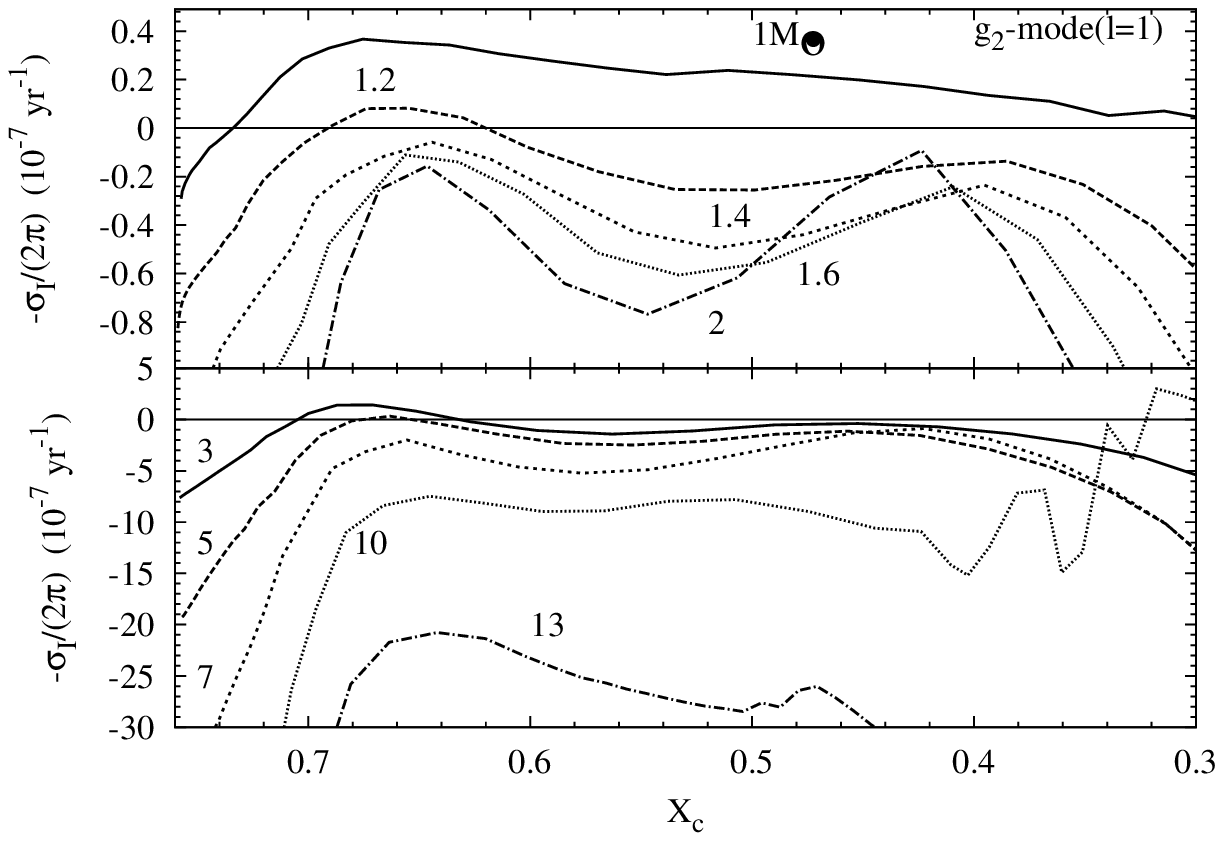}
  \end{center}
    \caption{Top: Variation of the growth rates of the dipole ($l=1$) g$_1$-mode with stellar evolution. The filled circles correspond to the equilibrium models of which the work integrals are shown in figure \ref{fig:04}. Bottom: Same as the top panel, but for the dipole g$_2$-mode.}
    \label{fig:05}
\end{figure}

Figure \ref{fig:04} shows the work integral for the dipole ${\rm g}_1$-modes for the models of 1, 1.2, 1.4, 1.6, 2, 3, 5, 7, 10, 13, 16, and 20\,$\Mo$ stars with the initial hydrogen abundance $X_0=0.76$ and $Z=0$. Each model corresponds to the most unstable (or the least stable) model along the evolutionary stage at around $X_{\rm c}\simeq 0.6$, which is marked as a filled circle in the top panel of figure \ref{fig:05}. As can be clearly seen in this figure, the nuclear burning core contributes to destabilization through the $\varepsilon$-mechanism, while the bulk of the outer part, except for the zone around $\log T\simeq 5.2$ in the case of $M\lesssim 2\,\Mo$, works against it. 
The positive contribution from $\log T\simeq 5.2$ is the $\kappa$-mechanism of the the second helium ionization zone, which is discussed later in detail.

As can be seen in figure \ref{fig:03}, the main nuclear source of stars with $M\lesssim 13\,\Mo$ around at the evolutionary phase of $X_{\rm c}\simeq 0.6$, which are shown to be vibrationaly unstable, is the pp-chain. On the other hand, in the case of more massive stars, $M \gtrsim 16\,\Mo$, which are shown to be vibrationaly stable, the main nuclear source is the CNO-cycle. As a consequence, we reach the conclusion that the population III stars with $M\lesssim 13\,\Mo$ are unstable against the dipole ${\rm g}_1$-mode, due to the $\varepsilon$-mechanism of the pp-chain. 

Figure \ref{fig:05} shows the growth rates of the dipole g$_1$- and g$_2$-modes as a function of the hydrogen mass function at the center, $X_{\rm c}$, which is an indicator of stellar evolution. The dipole g$_1$-mode becomes unstable from $X_{\rm c} \simeq$ 0.7 to 0.4 for $1.2\Mo\leq M \leq 10\,\Mo$. For $1\,\Mo$, instability grows rapidly after $X_{\rm c} \simeq 0.4$. We discuss it later in the next section. 

The growth rates of the g$_1$- and g$_2$-modes increase due to the $\varepsilon$-mechanism of the pp-chain with stellar evolution after leaving the ZAMS (figure \ref{fig:05}). As explained in Paper I, this delicate change of stability is caused by the variation in eigenfunction with evolution. A high-temperature sensitivity of the nuclear energy generation rate of the $^3{\rm He}$-$^3{\rm He}$ reaction is essential for the $\varepsilon$-mechanism to work in the case of the lower mass stars (\cite{Chris1974}). The effective value of $\varepsilon_T$ in the perturbation to the nuclear energy generation is raised to $\sim 11$ because of $^3$He($^3$He, 2$^1$H)$^4$He, while the value of $\varepsilon_T$ of the pp-reaction, itself, $^1$H($^1$H,e$^+ \nu_{\rm e}$)$^2$H, is $\simeq 4\sim 5$. Since the reaction $^3$He($^3$He, 2$^1$H)$^4$He occurs fast compared with the pp-reaction, $^3{\rm He}$ does not remain at the stellar center. On the other hand, the $^3{\rm He}$-$^3{\rm He}$ reaction does not efficiently occur with a decrease in temperature. As a consequence, $^3{\rm He}$ accumulates in a shell just outside the stellar center. Hence the most favorable situation for the vibrational instability is that the eigenfunction has a large amplitude in such an off-centered $^3{\rm He}$ shell. At the ZAMS stage, the accumulation of $^3{\rm He}$ is not enough, and thus the star is stable, or marginal, against any modes. As the evolution proceeds, the peak of the eigenfunction becomes to match with the off-centered $^3{\rm He}$ shell, and then the star becomes vibrationally unstable against the low-order g-modes.
For the higher mass stars, on the other hand, the pp-II and III branches contribute to the nuclear energy generation due to high temperature. The collisional reactions in these branches have high temperature dependence comparable with that of the $^3{\rm He}$-$^3{\rm He}$ reaction, and contribute to the g-mode instability.

In the case of stars with $M > 13\,\Mo$, the CNO-cycle is activated to work as the dominant source of the nuclear reaction at the evolutionary stage, $X_{\rm c} \simeq 0.7$. This is the reason why the unstable duration for the star with $M = 13\,\Mo$ is much shorter than for the less-massive stars, and why the more-massive stars do not have instability due to the $\varepsilon$-mechanism (see the top panel of figure \ref{fig:05}). 

In the case of a $10\,\Mo$ star, after the phase of $X_{\rm c} \simeq 0.4$, the growth rate changes rapidly, and it grows to be positive in the case of the g$_2$-mode. This is because the CNO-cycle becomes the dominant source of energy generation, and as a consequence the convective core grows without vanishing. A superadiabatic zone then appears outside the convective core, and some g-modes become excited by Kato's mechanism \citep{Kato1966}. We will discuss this instability and the relevant physics in detail in the forthcoming paper (in preparation).

\section{The $\kappa$-mechanism as a booster for instability}
It is well known that the $\kappa$-mechanism in the hydrogen and helium ionization zones is responsible for the Cepheid strip of radial pulsations. The depth of the helium ionization zone in stars having $\log T_{\rm eff} = 3.8\sim 4.0$ is appropriate to excite pulsation modes, of which frequencies are on the order of $(GM/R^3)^{1/2}$ by the $\kappa$-mechanism. The $\delta$ Sct stars, which are around $M\simeq 2\,\Mo$, are such pulsating stars near to the main sequence in the case of population I stars. 

Since the opacity is much lower in the metal-free population III stars than in the case of population I stars, the luminosity of population III stars is higher than that of population I stars of the same mass. Also, the population III stars are more compact than the population I stars. As a consequence, the effective temperature of population III stars is substantially higher than the population I stars. As cam be seen in figure \ref{fig:02}, in the case of metal-free population III stars, the main-sequence stars of $M\simeq 1\,\Mo$ is around $\log T_{\rm eff} = 3.8 \sim 4$, and hence the $\kappa$-mechanism of the the second helium ionization zone does work to excite low-degree, low-order p-modes in these stars.  

\begin{figure}[th]
  \begin{center}
 \FigureFile(80mm,80mm){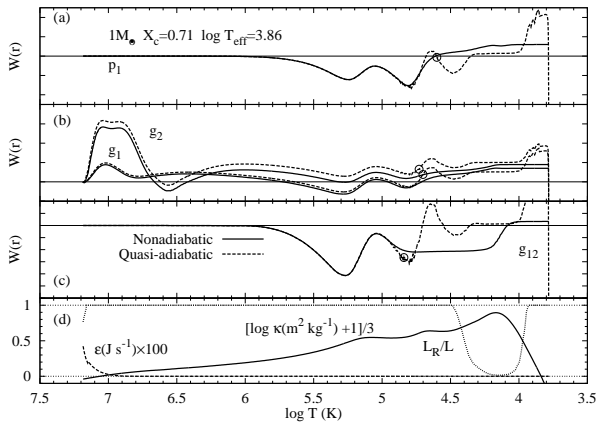}
  \end{center}
    \caption{(a): Work integral for the dipole p$_1$ mode of a $1\,\Mo$ star model at the evolutionary stage of $X_{\rm c} = 0.71$, $\log T_{\rm eff} = 3.86$. The solid line shows the results obtained by the fully nonadiabatic calculation, while the dashed line shows that obtained by the quasi-adiabatic approximation. The circle on the dashed curve for the quasi-adiabatic approximation indicates the location where the oscillation period is equal to the local thermal time scale.
(b): Same as (a), but for the dipole g$_1$- and g$_2$-modes. 
(c): Same as (a), but for the dipole g$_{12}$ mode.
(d): Opacity, $\kappa$, the nuclear reaction rate, $\varepsilon$, and the fraction of the radiative luminosity to the total luminosity, $L_{\rm R}/L$, as functions of $\log T$. }
    \label{fig:06}
\end{figure}

Panel (a) of figure \ref{fig:06} shows the work integral for the dipole p$_1$-mode. As can be seen in this figure, the p-modes are excited by the $\kappa$-mechanism, mainly due to the first helium ionization. Many of the other higher order p-modes and higher degree p-modes are also unstable. This situation is analogous to the situation of population I stars in the classical instability strip (\cite{Shibahashi1981}). It should be noted that, since the p-modes oscillate mainly in the outer envelope, where the density is much lower than in the deep interior, their amplitude grows 
more easily than g-modes, which oscillate in the deep interior. Hence, the growth rate of the modes excited by the $\kappa$-mechanism is generally  larger than those excited by the $\varepsilon$-mechanism. 

The positive contribution from $\log T\simeq 5.2$ seen in figure \ref{fig:04} in the case of $M \lesssim 2\,\Mo$ is the $\kappa$-mechanism of the the second helium ionization zone. Panel (b) of figure \ref{fig:06} shows the work integrals of the dipole g$_1$- and g$_2$-modes for a $1\,\Mo$ star model with $\log T_{\rm eff}=3.86$. This demonstrates that the $\kappa$-mechanism boosts up the work integral at the second helium ionization zone at around $\log T\simeq 5.2$,  and the first helium ionization zone at around $\log T\simeq 4.7$ after the contribution of $\varepsilon$-mechanism is almost cancelled by the radiative dissipation outside the nuclear burning core. It should be reminded here that in the case of the quasi-adiabatic approximation we cannot properly evaluate the contribution of the very outer envelope where $\log T \lesssim 4.7$  to the work integral, and we have to terminate the integration around the zone where the local thermal timescale becomes as short as the oscillation period. The current fully nonadiabatic calculation shows that the heat capacity above such zones is so small that the contribution of the first helium ionization zone and the hydrogen ionization zone are negligibly small. 

As the evolution proceeds, the Brunt-V\"ais\"al\"a frequency in the stellar interior increases due to an increase of the central condensation and the formation of the gradient of chemical compositions as a consequence of nuclear reactions converting hydrogen to helium, and hence the frequencies of g-modes increase. When the frequency of a g-mode is close to a p-mode, these two modes become to have a dual character: behaving with a g-mode character in the deep interior while with a p-mode character in the envelope. These eigenmodes never degenerate, however, and avoided crossings of modes occur, ---that is, the wave characters are switched for each other (see \cite{Unno1989}). 

\begin{figure}[th]
  \begin{center}
    \FigureFile(90mm,90mm){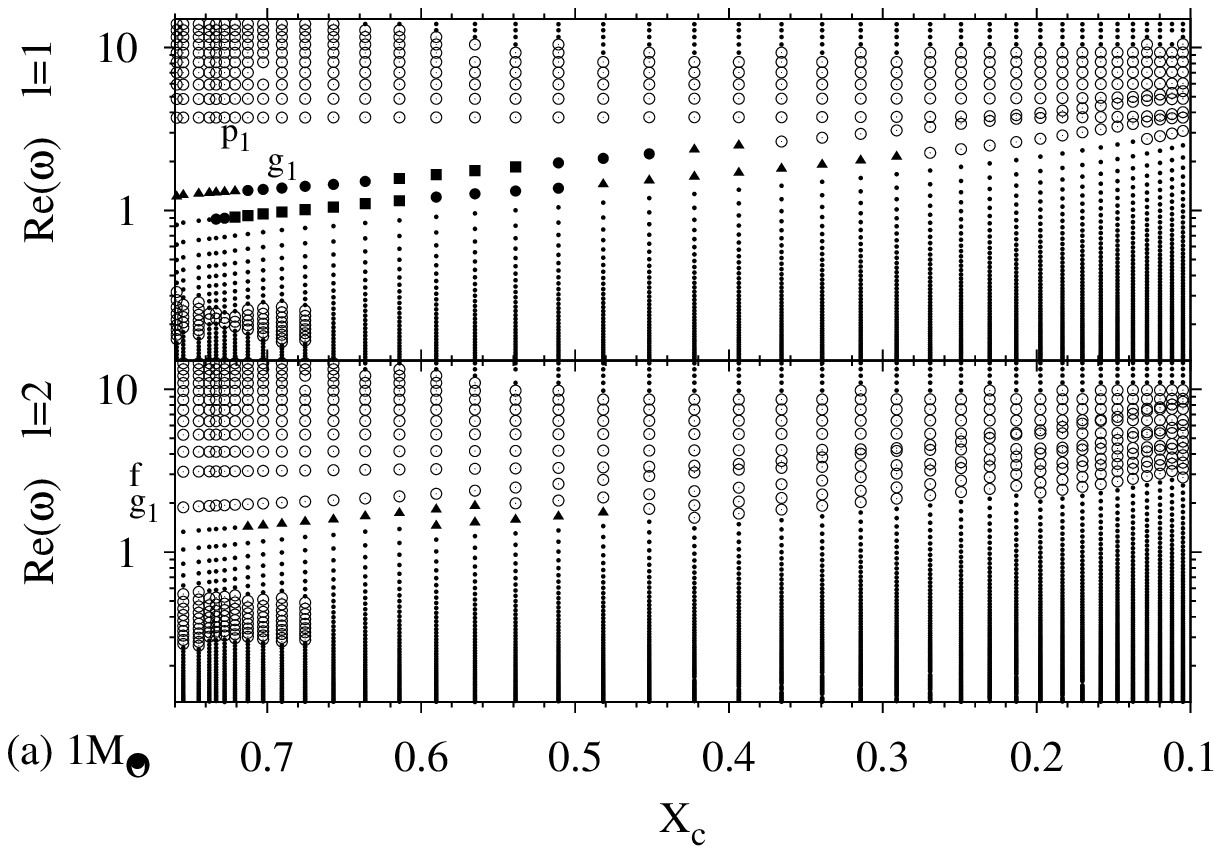}
    \FigureFile(90mm,90mm){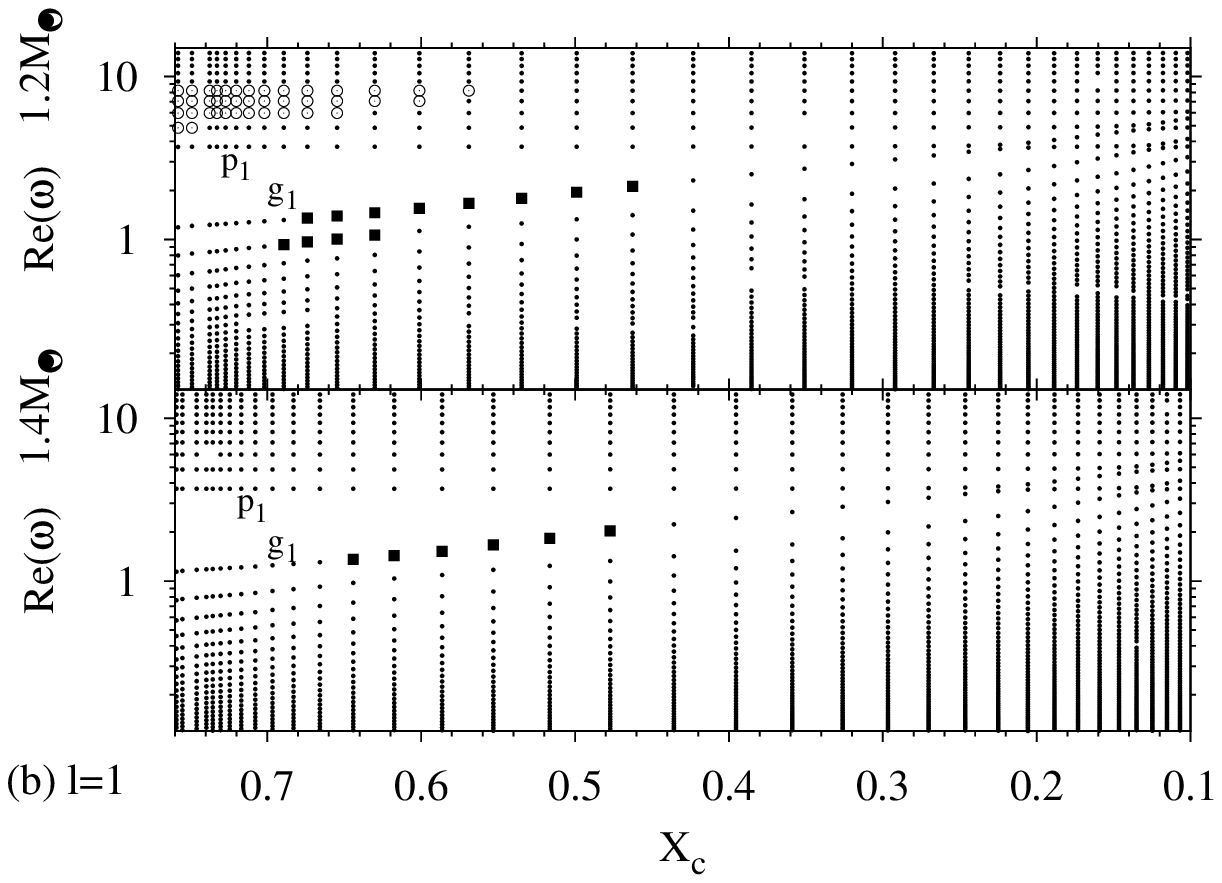}
  \end{center}
    \caption{(a): Variation in the frequencies and stability nature of the dipole ($l=1$) (upper panel) and quadrupole ($l=2$) (lower panel) modes for the metal-free $1\,\Mo$ star with evolution. The small dots denote stable modes, while the bigger symbols denote unstable modes. The latter are categorized into four types: (i) the modes that contribution of the $\varepsilon$-mechanism to the work integral is more than 50\%, shown with the filled squares, (ii) those that contribution of the $\varepsilon$-mechanism is 30-50\%, shown with the filled circles, (iii) those that contribution of the $\varepsilon$-mechanism is 10-30\%, shown with the filled triangles, and (iv) those excited essentially by the $\kappa$-mechanism, shown with the open circles. The abscissa is the central hydrogen abundance $X_{\rm c}$, which is an indicator of stellar evolution.
    (b): Same as (a), but for the dipole modes for the $1.2\,\Mo$ star (upper panel) and for the $1.4\,\Mo$ star (lower panel). }
    \label{fig:07}
\end{figure}

Figure \ref{fig:07} (a) shows the evolutionary variation in the nondimensional frequency, $\omega_{\rm R} \equiv\sigma_{\rm R} /(GM/R^3)^{1/2}$, normalized with the dynamical time scale, and the stability of the dipole and quadrupole modes of the $1\,\Mo$ star. In this diagram, the small dots denote the stable modes. As for the unstable modes, we classify them into four types by the degree of contribution of the $\varepsilon$-mechanism to the work integral: 
(i) the modes where the $\varepsilon$-mechanism contributes more than $50\,\%$,  
shown with filled squares, 
(ii) those with $30-50\,\%$,
shown with filled circles, 
(iii) those with $10-30\,\%$, 
shown with filled triangles, and 
(iv) those with less than $10\,\%$, 
shown by open circles. 
The modes of the type (i) can be regarded as being excited mainly by the $\varepsilon$-mechanism, while the type (iv) are regarded as being excited essentially by the $\kappa$-mechanism.

As can be seen in this figure, the p-modes of the $1\,\Mo$ star are vibrationally unstable during the main sequence stage. This is because the effective temperature of the metal-free $1\,\Mo$ star is high enough to be in the range of the instability strip on the HR diagram. Those p-modes are excited by the $\kappa$-mechanism working in the second helium ionization zone. 
With stellar evolution, the frequency of the g$_1$-mode approaches that of the p$_1$-mode, and the g$_1$-mode gradually becomes to have a mixed character of a p-mode and a g-mode. As a consequence, the relative contribution of the $\kappa$-mechanism to the work integral becomes larger and larger than that of the $\varepsilon$-mechanism. Hence, the growth rate of the g$_1$-mode becomes significantly larger. This is the reason why the growth rate of the dipole g$_1$-mode significantly increases.

With the increase of the stellar mass, the effective temperature at the main-sequence stage becomes higher. The $\kappa$-mechanism then becomes less efficient to excite oscillations. 
\citet{Marigo2001} showed that the instability strips for the radial fundamental mode, first and second overtones cross the main-sequence track for the population III $\sim 1\Mo$ star. Figure \ref{fig:07} (b) shows that in the case of a $1.2\,\Mo$ star, the dipole p$_1$-mode, of which the characteristics are similar to the radial fundamental mode, is stable during the hydrogen-burning phase, and that also the other p-modes are stable, except at the very early phase of evolution. In the case of $1.4\Mo$, the p-modes are no longer excited.

\section{Stability of low-degree, high-order g-modes}
It is found the $\kappa$-mechanism does work to excite low-degree high-order g-modes as well as p-modes simultaneously during the early phase of evolution. Panel (c) of figure \ref{fig:06} shows the work integral for the dipole g$_{12}$-mode. The amplitude of such a high-order g mode in the $^3{\rm He}$ burning shell is relatively much lower than in the outer envelope, and thus the contribution of the $\varepsilon$-mechanism to the stability of this mode is negligibly small. Rather, this mode is excited by the $\kappa$-mechanism of hydrogen ionization. The stability is, however, marginal. 

One might be reminded of hybrid population I pulsators of $M\simeq 1.5\,\Mo$, expected theoretically in relation to $\delta$\,Sct and $\gamma$\,Dor variables.
\citet{Guzik2000} first proposed ``convective blocking'' as the driving mechanism of high-order g-modes in the $\gamma$\,Dor variables. Dupret et al. (2004, 2005) performed a stability analysis while adopting the time-dependent convection theory. They found that this mechanism is dominant for the excitation, and hence that perturbation of convective energy flux has no significant effect. 
However, the situation is substantially different. In the present case of population III stars, the convective layer is induced only by hydrogen ionization, and it is much thinner than the population I stars; also, the high-order g-modes are excited by the $\kappa$-mechanism near to the top of the convective layer. Since the convective timescale there is shorter than the pulsation timescale, it is still hard to conclude definitely the stability of such case.

\section{Summary and discussion}
We performed a nonadiabatic analysis of the vibrational instability of population III stars due to the $\varepsilon$-mechanism of the pp-chain, which was analyzed with a quasi-adiabatic approximation in Paper I for stars $M<5\,\Mo$. We extended the mass range by adopting a stellar evolution code, MESA, which treats more appropriately the nuclear reaction networks after the onset of the CNO-cycle. We found that the dipole g$_1$-mode becomes unstable in a wider range of masses than investigated in Paper I, and that the stars of $M \lesssim 13\,\Mo$ become unstable at their early evolutionary phase, in which the pp-chain is the dominant energy source.
It was found that the dipole g$_2$-mode also becomes vibrationally unstable for some low-mass stars. 
This mode, however, is found to be unstable for a quite short fraction of the star's lifetime.
This mode becomes unstable for stars of $M \leq 1.2\Mo$ and $3\Mo \leq M \leq 5\Mo$. 

The $e$-folding time of the instability of the dipole g$_1$-mode is much shorter than the evolutionary timescale. Hence, it seems to be natural to expect that the instability grows to a finite amplitude. If the oscillatory wave grows to a huge amplitude, it is expected to induce material mixing, and to have a significant influence on the later evolution of the star, as was expected concerning the Sun by \citet{Dilke1972}. 

One of the possible influences on the stellar evolution might be an extension of the lifetime. That is, incorporation the surrounding cooler and hydrogen-rich matter into the central region could delay the hydrogen exhaustion.
It has already been expected that the lifetimes of the population III stars with $< 0.8\Mo$ should be longer than the age of the Universe, and should still be alive if ever born \citep{Marigo2001}.
If material mixing could be induced by the instability, the mass limit for the population III stars being still alive would be increased.
Hence, a possibility of detecting low-mass population III stars might be higher than expected before, and  there might remain a possibility of finding oscillations in such stars by monitoring a micro-magnitude level luminosity variation.  

Population III stars with $M\simeq 1\,\Mo$ were found to be unstable for both of the dipole g-modes and the low-degree p-modes simultaneously. The g-modes are excited by the $\varepsilon$-mechanism of the $^3$He-$^3$He reaction, and the p-modes are excited by the $\kappa$-mechanism in the helium ionization zone. This is a quite unique hybrid pulsator, only expected in metal-free population III stars. In the case of population I stars, the main-sequence stars in the classical instability strip are $\sim 2\,\Mo$ stars, in which the nuclear energy source is the CNO cycle, and  their g-modes are not excited by the $\varepsilon$-mechanism. At a certain evolutionary phase, avoided crossing occurs between the g-modes and p-modes. In such a case, the $e$-folding time of the g-modes significantly increases with the help of the $\kappa$-mechanism and becomes substantially shorter than the evolution timescale.    

This study has concentrated on the main-sequence stars.
It should be noted here that the number of known red giants with extremely meta-poor abundances is larger than that of dwarfs by a factor of two \citep{Suda2008}. Also, it should be reminded that solar-like p-mode oscillations in red giants with normal abundances are being detected in tens of thousands of stars with space-based photometric campaigns. Hence, one might be able to expect detection of solar-like p-mode oscillations in population III giants in the future. 
It should be remarked here that the extent of convective envelope might be different between population I stars and population III stars due to difference in the opacity. Thus, the convection spectrum, and hence property of stochastic excitation, are also expected to be different.

\bigskip
This study has been financed by Global COE Program ``the Physical Sciences Frontier'', MEXT, Japan. It adopted the MESA code to construct the stellar evolutionary models. The authors wish to acknowledge the \citet{Paxton2011} publication and MESA website\footnote{(http://www.mesa.sourceforge.net)} for providing their very useful evolution code to public as an open source.

\appendix
\section*{Evaluation of the density and temperature dependence of nuclear energy generation by the CN-cycle}

\begin{table}[h]
  \caption{The labels for the elements in the CN-cycle.
  \label{table:CN}}
  \begin{center}
    \begin{tabular}{ccc}\hline\hline
      $i$ & species & reaction \\ \hline
      1&$^{12}$C&$^{12}$C(p,$\gamma)^{13}$N   \\
      2&$^{13}$N&$^{13}$N(${\rm e}^-,\nu_{\rm e})^{13}$C    \\
      3&$^{13}$C&$^{13}$C(p,$\gamma$)$^{14}$N \\
      4&$^{14}$N&$^{14}$N(p,$\gamma)^{15}$O   \\
      5&$^{15}$O&$^{15}$O(${\rm e}^-,\nu_{\rm e})^{15}$N    \\
      6&$^{15}$N&$^{15}$N(p,$^4$He)$^{12}$C \\ \hline
    \end{tabular}
  \end{center}
\end{table}

This section introduces how to evaluate the density and temperature dependence of the nuclear energy generation by the CN-cycle, $\varepsilon_{\rho,{\rm CN}}$ and $\varepsilon_{T,{\rm CN}}$. We slightly modified the method shown in \citet{Kawaler1988} to obtain a reasonable result applicable even in $\log T \gtrsim 8$. 

If we label the species of the CN-cycle and proton as $i$'s, following table \ref{table:CN} and $p$, respectively, the differential equations governing the abundances of the reactants in the CN-cycle are written as follows:
\begin{eqnarray}
  N\frac{dy_i}{dt}&=& - N^2 y_i y_p C_i + N^2 y_{j} y_p C_{j}, \;\;\;(i=1,4) \label{eq:dy14}\\
  N\frac{dy_i}{dt}&=& - N y_i \lambda_i + N^2 y_{i-1} y_p C_{i-1}, \;\;\;(i=2,5) \label{eq:dy25}\\
  N\frac{dy_i}{dt}&=& - N^2 y_i y_p C_i + N y_{i-1} \lambda_{i-1}, \;\;\;(i=3,6) \label{eq:dy36}
\end{eqnarray}
where $N$ denotes the total number density and $y_i$ the number fraction of the element $i$;  $C_i$ denotes the rate of collisional destruction of species $i$ per proton for $i=1, 3, 4, 6$; and $\lambda_i$ is the rate of decay of species $i$ for $i=2, 5$. Here, the label $j$ means $i-1$, except for the case of $i=1$, in which the label $j$ on the right-hand-side of equation (\ref{eq:dy14}) should be recognized as $^{15}{\rm N}$ ($i=6$).
The Lagrangian variation of $y_i$'s is expressed as
\begin{equation}
  \frac{\delta y_i}{y_i}=\alpha_i \frac{\delta\rho}{\rho} + \beta_i \frac{\delta T}{T}.
  \label{eq:delta_y}
\end{equation}
That is, the coefficients $\alpha_i$'s and $\beta_i$'s are the density and temperature dependence of $y_i$, respectively.
By using these coefficients, the temperature and density dependences of the CN-cycle are written as
\begin{eqnarray}
  \varepsilon_{\rho,\mathrm{CN}}&=&\left.\left(\varepsilon_1+\varepsilon_3+\varepsilon_4+\varepsilon_6+\sum_i\alpha_i\varepsilon_i\right)\right/\varepsilon_{\mathrm{CN}}, \label{eq:eps_rho}\\
  \varepsilon_{T,\mathrm{CN}}&=&\left.\left(\nu_1\varepsilon_1+\nu_3\varepsilon_3+\nu_4\varepsilon_4+\nu_6\varepsilon_6+\sum_i\beta_i\varepsilon_i\right)\right/\varepsilon_{\mathrm{CN}}. \nonumber \\
  \label{eq:eps_t}
\end{eqnarray}
The reciprocal of the time scale for collisional destruction of species $i$ is
\begin{equation}
  K_i\equiv Ny_p C_i.
\end{equation}
Since the $C_i$'s depend only on temperature, and $N$ can be expressed as a function of the mass density, $\rho$, the Lagrangian variations of $K_i$'s are expressed as
\begin{equation}
  \frac{\delta K_i}{K_i}=\frac{\delta\rho}{\rho}+\nu_i\frac{\delta T}{T},
\end{equation}
where $\nu_i\equiv(\partial\ln C_i/\partial\ln T)_{\rho,y}$ is the temperature dependence of $C_i$.
Taking the Lagrangian variation of equations (\ref{eq:dy14})-(\ref{eq:dy36}), we have
\begin{eqnarray}
  \frac{\delta y_i}{y_i}&=&\frac{K_i}{i\sigma+K_i}\left[\alpha_{j}\frac{\delta\rho}{\rho}+(\beta_{j}+\nu_{j}-\nu_i)\frac{\delta T}{T}\right] \nonumber \\
    && \hspace{14em}(i=1,4) \\
  \frac{\delta y_i}{y_i}&=&\frac{\lambda_i}{i\sigma+\lambda_i}\left[(\alpha_{i-1}+1)\frac{\delta\rho}{\rho}+(\beta_{i-1}+\nu_{i-1})\frac{\delta T}{T}\right] \nonumber \\
 && \hspace{14em}(i=2,5) \\
  \frac{\delta y_i}{y_i}&=&\frac{K_i}{i\sigma+K_i}\left[(\alpha_{i-1}-1)\frac{\delta\rho}{\rho}+(\beta_{i-1}-\nu_i)\frac{\delta T}{T}\right] \nonumber \\
 && \hspace{14em}(i=3,6)
\end{eqnarray}

\begin{figure}[t]
  \begin{center}
    \FigureFile(90mm,90mm){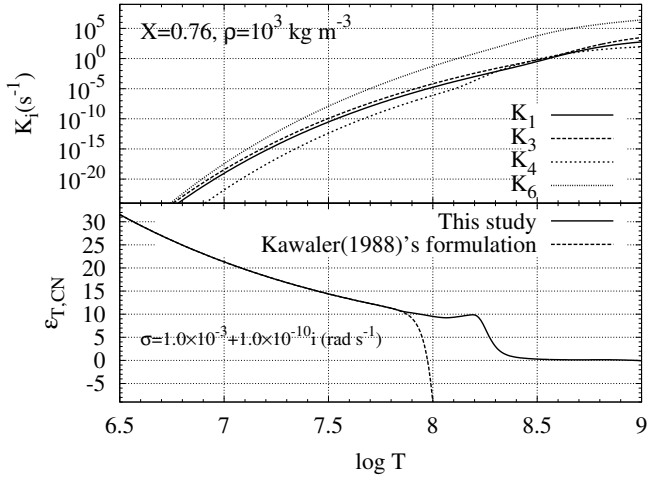}
  \end{center}
    \caption{Upper: Reciprocal of the timescale for collisional destruction of $^{12}{\rm C}$, $^{13}{\rm C}$, $^{14}{\rm N}$, and $^{15}{\rm N}$. 
    Lower: Temperature dependence of nuclear energy generation by the CN-cycle. For a comparison, the result by \citet{Kawaler1988} is shown with the dotted line. }
    \label{fig:appendix}
\end{figure}

Then, we obtain the recursive formulae for the $\alpha_i$'s:
\begin{eqnarray}
\alpha_1=\frac{K_1}{i\sigma+K_1}\alpha_6,\;\;\;\alpha_2=\frac{\lambda_2}{i\sigma+\lambda_2}(\alpha_1+1), \label{eq:alpha12} \\
\alpha_3=\frac{K_3}{i\sigma+K_3}(\alpha_2-1),\;\;\;\alpha_4=\frac{K_4}{i\sigma+K_4}\alpha_3,  \\
\alpha_5=\frac{\lambda_5}{i\sigma+\lambda_5}(\alpha_4+1),\;\;\;\alpha_6=\frac{K_6}{i\sigma+K_6}(\alpha_5-1) 
\end{eqnarray}
and for the $\beta_i$'s:
\begin{eqnarray}
  \beta_1=\frac{K_1}{i\sigma+K_1}(\beta_6+\nu_6-\nu_1),\;\;\;\beta_2=\frac{\lambda_2}{i\sigma+\lambda_2}(\beta_1+\nu_1),  \nonumber \\
  \label{eq:beta12} \\
  \beta_3=\frac{K_3}{i\sigma+K_3}(\beta_2-\nu_3),\;\;\;\beta_4=\frac{K_4}{i\sigma+K_4}(\beta_3+\nu_3-\nu_4), \nonumber \\
  \label{eq:beta34} \\
  \beta_5=\frac{\lambda_5}{i\sigma+\lambda_5}(\beta_4+\nu_4),\;\;\;\beta_6=\frac{K_6}{i\sigma+K_6}(\beta_5-\nu_6). \nonumber \\ 
  \label{eq:beta56}
\end{eqnarray}
Evaluating the $\alpha_i$'s and $\beta_i$'s by equations (\ref{eq:alpha12})-(\ref{eq:beta56}), we obtain $\varepsilon_{\rho,{\rm CN}}$ and $\varepsilon_{T,{\rm CN}}$ from equations (\ref{eq:eps_rho}) and (\ref{eq:eps_t}), respectively.


\begin{thebibliography}{}
\bibitem[Boury et al.(1975)]{Boury1975}
  Boury, A., Gabriel, A., Noels, A., Scuflaire, R., \& Ledoux, P. 1975, \aap, 41, 279
\bibitem[Boury \& Noels(1973)]{Boury1973}
  Boury, A., \& Noels, A. 1973, \aap, 24, 255
\bibitem[Christensen-Dalsgaard et al.(1974)]{Chris1974}
  Chiristensen-Dalsgaard, J., Dilke, F. W. W., \& Gough, D. O. 1974, \mnras, 169, 429
\bibitem[Dilke \& Gough(1972)]{Dilke1972}
  Dilke, F. W. W., \& Gough, D. O. 1972, Nature, 240, 262
\bibitem[Dupret et al.(2004)]{Dupret2004}
  Dupret, M.-A., Grigahc\`ene, A., Garrido, R., Gabriel, M., \& Scuflaire, R. 2004, \aap, 414, L17
\bibitem[Dupret et al.(2005)]{Dupret2005}
  Dupret, M.-A., Grigahc\`ene, A., Garrido, R., Gabriel, M., \& Scuflaire, R. 2005, \aap, 435, 927
\bibitem[Guzik et al.(2000)]{Guzik2000}
  Guzik, J. A., Kaye, A. B., Bradley, P. A., Cox, A. N., \& Neuforge, C. 2000, \apj, 542, L57 
\bibitem[Kato(1966)]{Kato1966}
  Kato, S. 1966, \pasj, 18, 374
\bibitem[Kawaler(1988)]{Kawaler1988}
  Kawaler, S. D. 1988, \apj, 334, 220
\bibitem[Marigo et al.(2001)]{Marigo2001}
  Marigo, P., Girardi, L.,  Chiosi, C., \& Wood, P. R. 2001, \aap, 371, 152 
\bibitem[Paxton et al.(2011)]{Paxton2011}
  Paxton, B., Bildsten, L., Dotter, A., Herwig, F., Lesaffre, P., \& Timmes, F. 2011, 
  \apjs, 192, 3
\bibitem[Shibahashi \& Osaki(1981)]{Shibahashi1981}
  Shibahashi, H., \& Osaki, Y. 1981, \pasj, 33, 427
\bibitem[Shibahashi et al.(1975)]{Shibahashi1975}
  Shibahashi, H., Osaki, Y., \& Unno, W. 1975, \pasj, 27, 401
\bibitem[Sonoi \& Shibahashi(2011)]{Sonoi2011}
  Sonoi, T., \& Shibahashi, H. 2011 \pasj, 63, 95 (Paper I)
\bibitem[Suda et al.(2008)]{Suda2008}
  Suda, T. et al. 2008, \pasj, 60, 1159
\bibitem[Suda et al.(2007)]{Suda2007}
  Suda, T., Fujimoto, M. Y., \& Itoh, N. 2007, \apj, 667, 1206
\bibitem[Unno(1975)]{Unno1975}
 Unno, W. 1975 \pasj, 27, 81
\bibitem[Unno et al.(1989)]{Unno1989}
  Unno, W., Osaki, Y., Ando, H., Saio, H., \& Shibahashi, H. 1989, Nonradial oscillations of stars (University of Tokyo Press, 1989, 2nd ed.) 
\bibitem[Weiss et al. (2000)]{Weiss2000}
 Weiss, A., Cassisi, S., Schlattl, H., \& Salaris, M. 2000, \apj, 533, 413
\end{thebibliography}
\end{document}